\documentclass{article}
\usepackage[english]{babel}
\usepackage[utf8]{inputenc}

\usepackage[letterpaper,top=2cm,bottom=2cm,left=2.5cm,right=2.5cm,marginparwidth=1.75cm]{geometry}

\usepackage[raggedright]{titlesec}

\usepackage{url}
\usepackage{amsmath}
\usepackage{amsfonts}
\usepackage{graphicx}
\usepackage{xcolor}
\usepackage{caption}
\usepackage{subcaption}
\usepackage{hyperref}
\usepackage{cleveref}
\usepackage{dirtytalk}
\usepackage{authblk}
\usepackage{algorithm}
\usepackage{algpseudocode}

\newcommand{\R}{\mathbb{R}}

\newcommand{\dd}{\mathrm{d}}

\newcommand{\x}{x}
\newcommand{\cv}{\psi}
\newcommand{\Fcv}{F}
\newcommand{\xc}{x_\perp}

\newcommand{\ts}{t}

\newcommand{\revised}[1]{{#1}}

\begin{document}
\title{Coarse Grained Molecular Dynamics with Normalizing Flows}% Force line breaks with \\

\author{Samuel Tamagnone$^{1}$, Alessandro Laio$^{\dagger,1,2}$,  Marylou Gabrié$^{*,3}$}
\date{\small
    $^1${International School for Advanced Studies (SISSA), Via Bonomea 265, Trieste, Italy}\\
    $^2${The Abdus Salam International Centre for Theoretical Physics (ICTP), Strada Costiera 11, Trieste, Italy}\\
    $^3${CMAP, CNRS, École polytechnique, Institut Polytechnique de Paris, 91120 Palaiseau, France}\\
    $^\dagger$ laio@sissa.it,
    $^*$ marylou.gabrie@polytechnique.edu, 
}

%  \altaffiliation[Also at ]{Physics Department, XYZ University.}%Lines break automatically or can be forced with \\
%  \email{Second.Author@institution.edu}
% \affiliation{%
% SISSA, École Polytechnique
% }%

% \date{\today}
\maketitle
\begin{abstract}
    
We propose a sampling algorithm relying on a collective variable (CV) of mid-size dimension modelled by a normalizing flow and using non-equilibrium dynamics to propose full configurational moves from the proposition of a refreshed value of the CV made by the flow. The algorithm takes the form of a Markov chain with non-local updates, allowing jumps through energy barriers across metastable states. The flow is trained throughout the algorithm to reproduce the free energy landscape of the CV. The output of the algorithm are a sample of thermalized configurations and the trained network that can be used to efficiently produce more configurations. We show the functioning of the algorithm first on a test case with a mixture of Gaussians. Then we successfully test it on a higher dimensional system consisting in a polymer in solution with a compact and an extended stable state separated by a high free energy barrier. 
\end{abstract}

\section{Introduction}

Sampling from the Boltzmann distribution is a fundamental task to understand the equilibrium properties of physical systems. For molecular systems, molecular dynamics (MD) simulations mostly relies on local-updates inspired by physical dynamics. Yet, in the presence of local free energy minima caused by energetic or entropic barriers, these simulations fail to sample from the Boltzmann distribution of metastable systems as it takes an exponential time for physical dynamics to escape such minima. 

To tackle this issue, one strategy is to resort to enhanced sampling techniques. Some of these approaches are based on the knowledge of a low-dimensional collective variable (CV) capturing the metastability which is used to drive the sampling between metastable states \cite{eric, tuckerman, wham1, transition_interface_sampling, umbrella, thermo, pohorille,LP02}. CV-based enhanced samplers have two requirements. First, that for a fixed value of the CV, the remaining degrees of freedom can be efficiently equilibrated. Second, that the CV space is small enough to be explored efficiently. Meeting these requirements simultaneously is not an easy task as restricting to a handful of dimension to summarize the metastability of the system may be very difficult, if not simply inadequate.  
\revised{A significant research effort has been devoted to the development of approaches which allow performing enhanced sampling in high-dimensional CV spaces. Metadynamics-based approaches allow estimating simultaneously arbitrarily many low-dimensional free energies \cite{piana2007bias,prakash2018biasing}, but not the free energy as a simultaneous function of many variables. In these approaches the free energy as a function of $\mathcal{O}(10)$ variables can be obtained by reweighting \cite{marinelli2009kinetic}. Temperature Accelerated Molecular Dynamics \cite{tamd,tamd1} allows biasing simultaneously tens or hundreds of collective variables, which are artificially kept at high temperature to beat metastability. However, also this approach does not provide an explicit estimate of the free energy in such a high dimension, but only samples harvested from the corresponding probability density.}

A different line of work recently leveraged a class of generative models called normalizing flows (NF) \cite{papamakariosNormalizingFlowsProbabilistic2021a} to enhance the sampling. There, NF are used to propose candidate configurations subsequently reweighed or incorporated into a Markov chain with an accept-reject step; see the pioneering works of \cite{wuSolvingStatisticalMechanics2019,albergoFlowbasedGenerativeModels2019,noeBoltzmannGeneratorsSampling2019}. As the NF generates independent configurations across the entire configurational space, NF-based samplers are agnostic to energetic and entropic barriers and therefore particularly advantageous for metastable systems. However, NFs were found to hit a limit in the dimension and complexity of systems they can handle \cite{deldebbioEfficientModellingTrivializing2021,mahmoudAccurateSamplingMacromolecular2022,ciarellaMachinelearningassistedMonteCarlo2023,greniouxSamplingApproximateTransport2023a}. 

In this work, we present a Markov chain Monte Carlo (MCMC) sampling algorithm combining the use of a (high-dimensional) CV and a NF.
At each iteration, a refreshed CV value is proposed by a NF and a new configuration is constructed given the refreshed CV. The new configuration is accepted or rejected following the framework of non-equilibrium candidate Monte Carlo \cite{athenesComputationChemicalPotential2002,nilmeierNonequilibriumCandidateMonteCarlo2011,chenGeneralizedMetropolisAcceptance2015}, which guarantees that the stationary measure of the defined Markov Chain is the target Boltzmann distribution. The NF is trained adaptively along the MCMC to approach the image of the Boltzmann measure through the CV-map related to the CV free energy. This learning procedure entirely avoids the requirement for a dataset of thermalized configurations beforehand and improves the NF proposals along the way.
By requiring the NF to learn a distribution in CV space rather than the configurational space, the method benefits from the exploration power of these generative models while easing its training by reducing dimension. The proposed method also revisits the idea of a CV-driven sampler while significantly relaxing the constraint on the dimension of the CV, precisely because the NF can drive jumps between metastable states in CV-space. Thus, the CV dimension must be smaller than the total number of degrees of freedom but can be significantly larger than what is allowed by a traditional CV-enhanced sampler. We refer to a regime of \emph{mid-size} for the dimension of CV handled by the proposed algorithm,
\revised{which can range from some units to a few thousands of dimensions depending on the system under study. As a typical example, one can think of a molecule in solution, in which the main degrees of freedom of the molecule can be selected as CVs. While it would be impossible for a NF to model accurately the full system of molecule and solvent, it was shown that a NF can learn the distribution of relatively large molecules, for instance a protein of 892 atoms (2676 dimensions) \cite{noeBoltzmannGeneratorsSampling2019}. In the proposed method, the NF notably learns explicitly the free energy of the tens-to-hundreds dimensional CV, which is not achieved by competing enhanced sampling method handling high-dimensional CVs.}

A popular choice of a mid-size CV for a molecule is a coarse-grained representation of the chain of atoms. Coarse-grained simulations of molecular systems have brought significant insights on large molecular systems which are unreachable by full-atomistic simulations, in particular for events happening on large-time scales  \cite{clementiCoarsegrainedModelsProtein2008,pakAdvancesCoarsegrainedModeling2018}. However, designing a statistically unbiased sampler for the coarsed-grained Boltzmann measure is highly non trivial as it relates to the CV free energy, which is computationally intractable due to high dimension of the CV. Using smart proposals of CV values therefore requires to `backmap' this coarsed-grained configuration to a full-atomistic configuration on which the fine-grained energy function can be evaluated so as to reweigh, or decide whether to accept, the proposal. This `backmapping' remains a significant challenge. A line of recent works explored the ability of deep generative models to help with this task \cite{liBackmappingCoarsegrainedMacromolecules2020,wangGenerativeCoarseGrainingMolecular2022,mahmoudAccurateSamplingMacromolecular2022,chennakesavaluEnsuringThermodynamicConsistency2023,chennakesavaluDataEfficientGenerationProtein2024}. Yet, backmapping strategies based on conditional NFs are eventually limited by their already mentioned restricted expressiveness \cite{mahmoudAccurateSamplingMacromolecular2022,chennakesavaluEnsuringThermodynamicConsistency2023}.
Conversely, methods involving other types of generative models are typically more accurate but are crucially lacking the tractable likelihood necessary to compute reweighing factors and acceptance probabilities. Here, we take a different route by using an NF generative model exclusively in the CV space and resorting to non-equilibrium dynamics to move the remaining degrees of freedom \cite{nilmeierNonequilibriumCandidateMonteCarlo2011, chenGeneralizedMetropolisAcceptance2015}. 

The article is organized as follows. In \cref{sec:algo-description} we detail the proposed algorithm. In \cref{sec:numerics}, we first demonstrate our method on a control example of mixture of Gaussians showing the robustness of the method to dimension and energy-barrier height. We also validate the method on a molecular system of a 9-beads polymer in a Lennard-Jones solvent, confirming here again the robustness to energy-barrier height.
\revised{While our results do not yet demonstrate the suitability of the method at the level of complexity of a protein folding in water, they show promise of the approach in handling realistic systems which shall be confirmed with more work, including interfacing a MD framework.}

\section{Algorithm description}
\label{sec:algo-description}

We are interested in sampling a physical system with configurations $\x \in \R^d$ harvested from a Boltzmann distribution with energy $U(x)$ and inverse temperature $\beta$:
\begin{align}\label{eq:boltzmann}
   \rho_*(x) = e^{-\beta U(x)} /  Z.
\end{align}
We assume that $\x$ can be decomposed into a (typically multidimensional) collective variable (CV) $\cv \in \R^n$ and transversal degrees of freedom $\xc \in \R^{d-n}$, that is $\x = (\psi, \xc)$.
We denote by $\Psi: \R^d \to \R^n$ the mapping $\x \to \Psi(\x)=\cv$.
The coordinate decomposition translates into a decomposition of the Boltzmann distribution:
\begin{align}
\rho_* (x) = \rho_*(\x|\cv) \rho^{\Psi}_*(\cv)
\end{align}
with 
\begin{align}\label{eq:fe}
\rho^{\Psi}_*(\cv) = \int_{\R^d} \rho_*(\x) \delta( \psi - \Psi(\x))\dd \x = e^{-\beta \Fcv_*(\cv)},
\end{align}
where $\rho_*(\x|\cv)$ is the density of the Boltzmann distribution \eqref{eq:boltzmann} conditioned on the CV value $\psi$ and we define the CV free energy $\Fcv_* : \R^n \to \R$.

The proposed algorithm combines ideas from the non equilibrium candidate Monte Carlo (NCMC) framework \cite{nilmeierNonequilibriumCandidateMonteCarlo2011} and from the adaptive Monte Carlo augmented with normalizing flows (NF), sometimes referred to as flowMC \cite{gabrieAdaptiveMonteCarlo2022}. The latter consists in an MCMC iterating three basics steps: Metropolis-adjusted Langevin dynamics (MALA updates), non-local moves proposed by a NF (non-local updates) and training steps of the NF to reproduce the distribution of observed MCMC samples (NF training). The procedure is called adaptive as the NF involved in the non-local update kernel learns from the previous samples produced by the MCMC. The method exploits the two specific characteristic of NF among the available generative models, they are easy to sample from and the density of the probability distribution they represent is easily evaluated \cite{papamakariosNormalizingFlowsProbabilistic2021a}. 

Here, we follow the three steps of flowMC but amend the algorithm by using a NF to propose non-local moves in a mid-sized CV space rather than in high-dimensional configuration-space. From a refreshed CV value, a move in configuration space is then proposed by steering the CV towards this value while relaxing accordingly the transversal degrees. The steered proposal is then accepted with a probability following the NCMC framework. The NF is trained to reproduce the statistics of the CV values visited by the Markov chain. The algorithm resulting from the iteration of MALA updates, steered updates, for multiple parallel walkers, and NF training is summarized in \cref{alg:CVampling_algo} and in \cref{fig:explicative_image_polymer}. We explain in detail the steered update and NF training below.

\paragraph{Steered update} The steered moves are decomposed in three steps. Assume we start from a configuration $x = (\cv, x_\perp) $. First the 
% , at the non-local step, 
NF proposes the new CV value $\cv'$ independently from $x$; the flow density is denoted by $\rho_\theta^{\Psi}(\cv)$ where $\theta$ stands for the trainable parameters. Second, we update the CV in such a way that in $N$ steps it reaches the target value. We present the algorithm assuming that this protocol is a simple linear interpolation, but our derivation is valid also for a generic protocol. In the linear interpolation case
\begin{align}
    \cv^{\ts+1} = \cv^\ts + \frac{\Delta\cv}{N}, 
\end{align}
where $\Delta\cv = \cv'-\cv$. In each step of the interpolation we  perform MALA relaxation steps for the remaining degrees of freedom, with proposals 
\begin{align}
    \widetilde x_\perp = x^\ts_\perp - \frac{\tau}{\gamma} \nabla_{x_\perp} U\left(\cv^\ts + \Delta\cv/2N, \, x^\ts_\perp\right) + \sqrt{\frac{2\tau}{\beta\gamma}}\eta 
\end{align}
where $\tau$ is a finite time step, $\gamma$ a friction coefficient, $\nabla_{x_\perp}$ the gradient operator at fixed CV and $\eta$ a sample from the standard normal distribution $\mathcal{N}(0,1)$.
If the proposal $\widetilde x_\perp$ is accepted -- according to the Metropolis-Hasting criteria for the conditional target $\rho_*(x_\perp|\cv^\ts)$ -- then $x_\perp^{\ts+1} = \widetilde x_\perp$, otherwise $x_\perp^{\ts+1} = x_\perp^\ts$; see \cref{alg:mala_algo} in the Supplementary Information for a reminder of the MALA algorithm. 
Third, the full move $(\cv, x_\perp) \rightarrow (\cv', x_\perp') = (\cv^N, x_\perp^N)$ is accepted with probability $p=\min[ \, 1 , \, R\,]$ with 
\begin{align}\label{eq:acc-prob}
R = \frac{\rho_\theta^\Psi(\cv)}{\rho_\theta^\Psi(\cv')}\exp (-\beta W). 
\end{align}
The first factor in $R$ accounts for a ratio of proposal probabilities, while the second depends on the work performed on the system during the steered move: 
\begin{align}
W= U(\cv', x_\perp')-U(\cv, x_\perp) + \sum_{\ts=0}^{N-1}\Big( U(\cv^\ts+\Delta\cv/2, x_\perp^\ts)-U(\cv^\ts+\Delta\cv/2, x_\perp^{\ts+1})\Big).
\end{align}

This procedure samples the Boltzmann distribution \eqref{eq:boltzmann} as the acceptance criterion satisfies detailed balance \cite{nilmeierNonequilibriumCandidateMonteCarlo2011,frenkelSpeed-upOfMonteCarlo2004}. A parallel work,  \cite{christoph} discusses further the conditions for the reversibility of a CV-steered update. 
In practice, the number of steps $N$ has implications on the acceptance probability and the computational cost of each steered update. It can be chosen in a system-dependent optimal way by running a preliminary study on a series of steered moves and choosing, as a function of the steering distance $\Vert\Delta\cv\Vert_2$ in CV space, the $N$ that yields the maximal $p/N$ ratio, with $p$ the acceptance probability \eqref{eq:acc-prob}. Intuitively, the larger the distance the more the transversal degrees of freedom are perturbed and the higher is the optimal $N$ to use. The approximate law for the optimal $N(\Vert\Delta\psi\Vert_2)$ is then used at run time.

\paragraph{NF training}
At each iteration, the NF parameters are updated to learn the marginal probability distribution on the collective variables. By analogy with flowMC \cite{gabrieAdaptiveMonteCarlo2022}, we seek to minimize the Kullback-Leibler divergence between $\rho_*^\Psi$ and $\rho_\theta^\Psi$,
\begin{align}
D_{\rm KL}(\rho_*^\Psi\mid\mid \rho_\theta^\Psi) =-\int_{\R^n}\log(\rho_\theta^\Psi(\psi))\rho_*^\Psi(\psi) d\psi + C_*
\end{align}
where $C_*$ is a constant independent of the parameters $\theta$ of the NF. For each gradient descent step, a batch of configurations $\{x^k\}_{k=1}^{n_{\rm batch}}$ is randomly selected from the states previously visited by the parallel walkers of the algorithm. The empirical loss function then is the negative log-likelihood in CV space
\begin{align}
L[\theta] = - \frac{1}{n}\sum_{i=k}^n\log(\rho_\theta^\Psi(\Psi(x^k))).
\end{align}
In the literature of NF-assisted samplers, this maximum likelihood training has been called ``training by example", as we use a set of states, which, at convergence, are Boltzmann-distributed, to compute the loss function. We do not perform ``training by energy'', where the reverse Kullback-Leibler divergence is minimized \cite{rezendeVariationalInferenceNormalizing2015,albergoFlowbasedGenerativeModels2019,noeBoltzmannGeneratorsSampling2019,wuSolvingStatisticalMechanics2019}. For a NF defining a density $\rho_\theta$ in full configuration space the reverse Kullback-Leibler divergence $D_{\rm KL}(\rho_\theta \mid\mid \rho_*)$ is an expectation on $\rho_\theta$ and can be estimated by a Monte Carlo average from the knowledge of the energy function $U$. Note that in our case this is unfeasible as it would require evaluating the intractable free energy $F_*(\cv)$ defined in \eqref{eq:fe}.

\begin{algorithm}
\caption{CV-space adaptive Monte Carlo}\label{alg:CVampling_algo}
\begin{algorithmic}
\Require  $n_{\rm iter}$ number of iterations of the algorithm, $\{x_i^0(0)\}_{i=1}^{n_{\rm walkers}}$ initial set of configurations, $\beta$ inverse temperature, $U(\cdot)$ energy function, $\Psi(\cdot)$ collective variable mapping,
$\theta$ initial NF parameters,
$\epsilon$ learning rate, $n_{\rm batch}$ batch-size,
$n_{\rm local}$ number of local steps, $N(\cdot)$ steering-schedule number-of-steps rule, $\tau$ time step for MALA, $\gamma$ friction coefficient.
\\

\State $\mathcal{B} = \{\}$ \Comment{initialize buffer}
\For{$n = 0, \dots, n_{\rm iter}$}
\For{$i = 1, \dots, n_{\rm walkers}$} 
\State $\{x^{i,\alpha}(n)\}^{\alpha=1, \dots, n_{\rm local}} = {\rm MALA}(U(\cdot), \beta, \gamma, \tau, x_i^0(n), n_{\rm local})$ \Comment{update the walkers with local moves}\footnotemark
\State $x^{i,0}(n+1) = {\rm STEERED}(U(\cdot), \beta, \gamma, \tau, x_i^{n_{\rm local}}(n), N(\cdot), \rho_\theta^\Psi(\cv))$ \Comment{use NF for a non-local move}
\State $\mathcal{B} \gets \mathcal{B} \cup \{x^{i,0}(n+1)$\}
\EndFor
\For{$k=1, \dots n_{\rm batch}$}
\State $x^k \sim \mathcal{U(\mathcal{B})}$ \Comment{sample training examples uniformly over the buffer}
\EndFor
\State $L[\theta] = \frac{-1}{n_{\rm batch}}\displaystyle\sum_{k=1}^{n_{\rm batch}}\log(\rho_\theta^\Psi(\Psi(x^k)))$ \Comment{compute negative log-likelihood 
}
\State $\theta \gets \theta - \epsilon \nabla_\theta L[\theta]$ \Comment{make one stochastic gradient descent update on the NF parameters}\footnotemark
\EndFor
\\
\Return $\theta$ trained NF parameters, $\{x^{i,\alpha}(n)\}_{i=1, \dots, n_{\rm walkers}}^{n=1, \dots, n_{\rm iter};\ \alpha = 0, \dots, n_{\rm local}}$ all visited configurations\footnotemark
\end{algorithmic}
\end{algorithm}
\footnotetext[2]{In SI \cref{alg:mala_algo,alg:steered_algo} the reader may find the precise description of the STEERED update and of the MALA update.}
\footnotetext[3]{One could decide to do more than one training step each non-local move, as this is the more computationally costly part of the simulation. This is what we have done in our numerical examples.}
\footnotetext[4]{Typically, one does not save the configurations reached at all the local steps, but only a sub-sample of those, as they can be strongly correlated.}

\section{Numerical results}
\label{sec:numerics}
\paragraph{Mixture of Gaussians}
As a first proof of concept, we test our procedure on mixture of Gaussian distributions. We show that the NF correctly learns the free energy of the system and that, at convergence, the average acceptance of the steered moves does not depend on the height of the free energy barrier  in CV space. As a consequence, if the free energy barrier depends only weakly on the transversal degrees of freedom, namely if the CV are well chosen, the cost of the simulation is approximately independent of the barrier height.

\begin{figure}
    \centering
    \includegraphics[width=\textwidth]{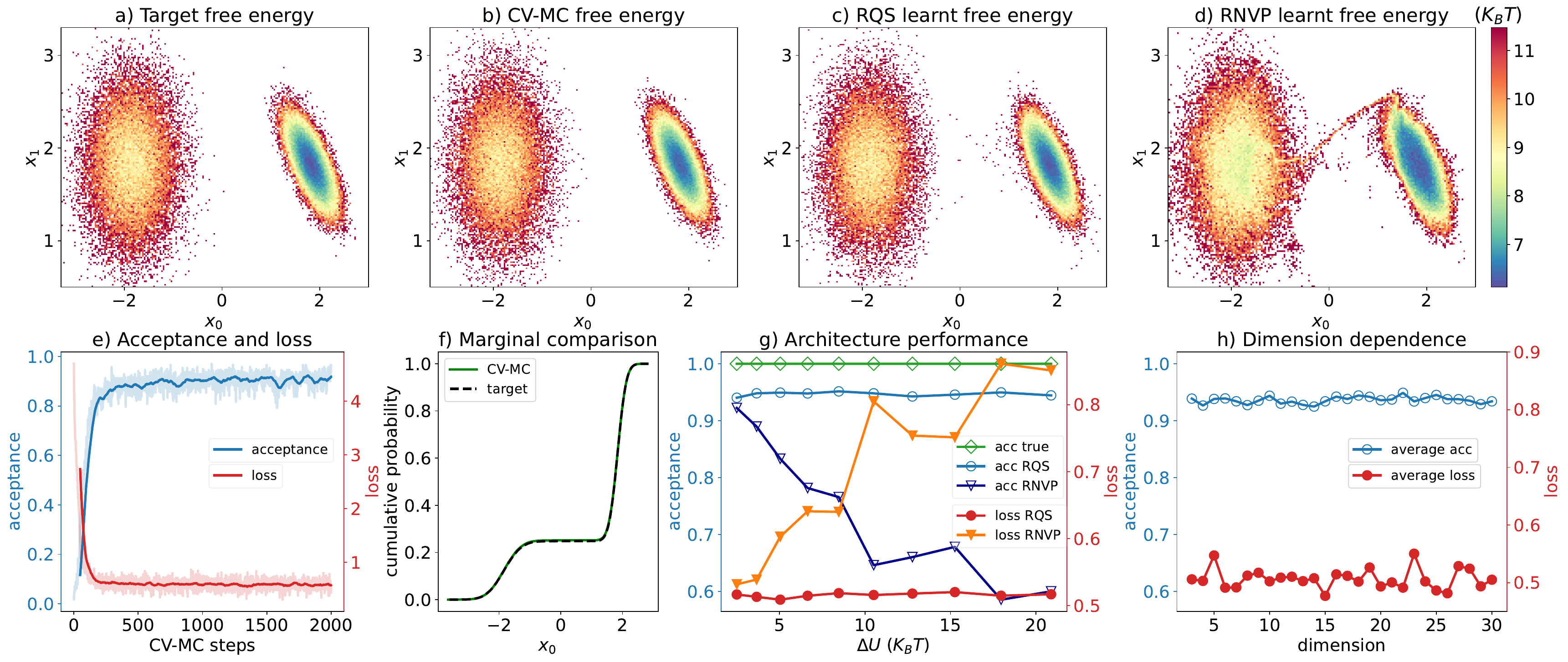}
    \caption{a), b), c) and d) maps of the true, output of the CV-MC, learnt by RQS or RNVP free energy landscapes. e) acceptance and loss function of a test case with the RQS NF architecture during the first 2k steps of the algorithm. f) comparison of the cumulative distribution on x for the target distribution and the CV-MC one for checking if the weights of the modes are correctly reproduced. g) performance in the learning task obtained with different architectures and different energy barriers. h) acceptance and loss at convergence for different numbers of transversal degrees of freedom.}
    \label{fig:gaussian_experiment}
\end{figure}

We first consider a mixture of two Gaussians in 3D. Here, the full states are described by vectors $x = (x_0, x_1, x_2) \in \R^3$ and the chosen CVs are the first two coordinates $x_0$ and $x_1$ ($\Psi(x) = (x_0, x_1)$), leaving one transversal degree of freedom $x_2$. The example is set up so that the CVs allow distinguishing between the two different modes (see Materials \& Methods below for details). The results are presented in \cref{fig:gaussian_experiment}. In panel a) we show the true 2D free energy of the mixture that we are reproducing with the algorithm. In b) we show the free energy obtained by the states saved alongside the CV-MC iterations. In panels c) and d) we show the free energy learnt along the proposed adaptive algorithm by two different NF architectures: the Rational Quadratic Spline (RQS) \cite{neuralsplineflows} in c) and the Real NVP (RNVP) \cite{realnvp} in d).  As shown in panels a), b), c) and d) there is a good agreement between the target free energy, the learnt one and the one computed with the output of the CV-space adaptive Monte Carlo. Because of its smoothness, the normalizing flow has a tendency to connect the two separated modes. This occurs especially for the RNVP, as shown in d), because it is less expressive than the RQS.
This small systematic error in the free energy learned by the flow is well corrected by the accept/reject step in the non-local updates. Indeed the CV of the samples produced by the algorithm are distributed correctly. In panel f), we further check that the cumulative distribution of the marginal law on $x_0$ for the target and simulated distribution (with the RQS) match almost perfectly. In panel e) we show the behavior of the acceptance and of the loss as a function of the algorithm's iterations (again using RQS). As the NF learns the underlying free energy, the algorithm becomes more and more efficient in making proposals and the acceptance grows. As the plot shows, 2000 training iterations are already enough to reach convergence.

We then consider the same mixture of two Gaussians example, but increase the free energy barrier by pulling further apart the mixture components. 
The plots in panel g) illustrate the most important advantage of the procedure:
the average acceptance at convergence depends very weakly on the free energy barrier. 
The small dependence is caused only by the slightly varying ability of the NF to learn the underlying  probability.
Indeed, when the proposal is the exact target marginal probability (green line), as if the NF has perfectly learnt the underlying marginal, the acceptance of NF-proposed moves is constant.
We have repeated the test with the two different NF architectures mentioned above: the RNVP, which is less expressive, and the more expressive RQS. 
The decrease in the acceptance that we have with the RNVP (dark blue line) follows exactly the increase in the loss (orange line), which means that this decrease is caused by proposals that become less and less accurate because the RNVP has more difficulties learning modes that are further apart, as it involves greater non-linearities. This does not occur with the RQS, which is more expressive. Hence we have decided to use that architecture in the tests on the polymer model presented below.

Finally, in panel h) we repeat the same numerical experiment adding more transversal degrees of freedom (up to 28), but keeping the same 2D free energy landscape. As shown by the average loss at convergence in CV space but also by the average acceptance in full space, the result is independent of the number of transversal dimensions if the CVs are properly selected.

\begin{figure}
    \centering
    \includegraphics[ width=\textwidth]{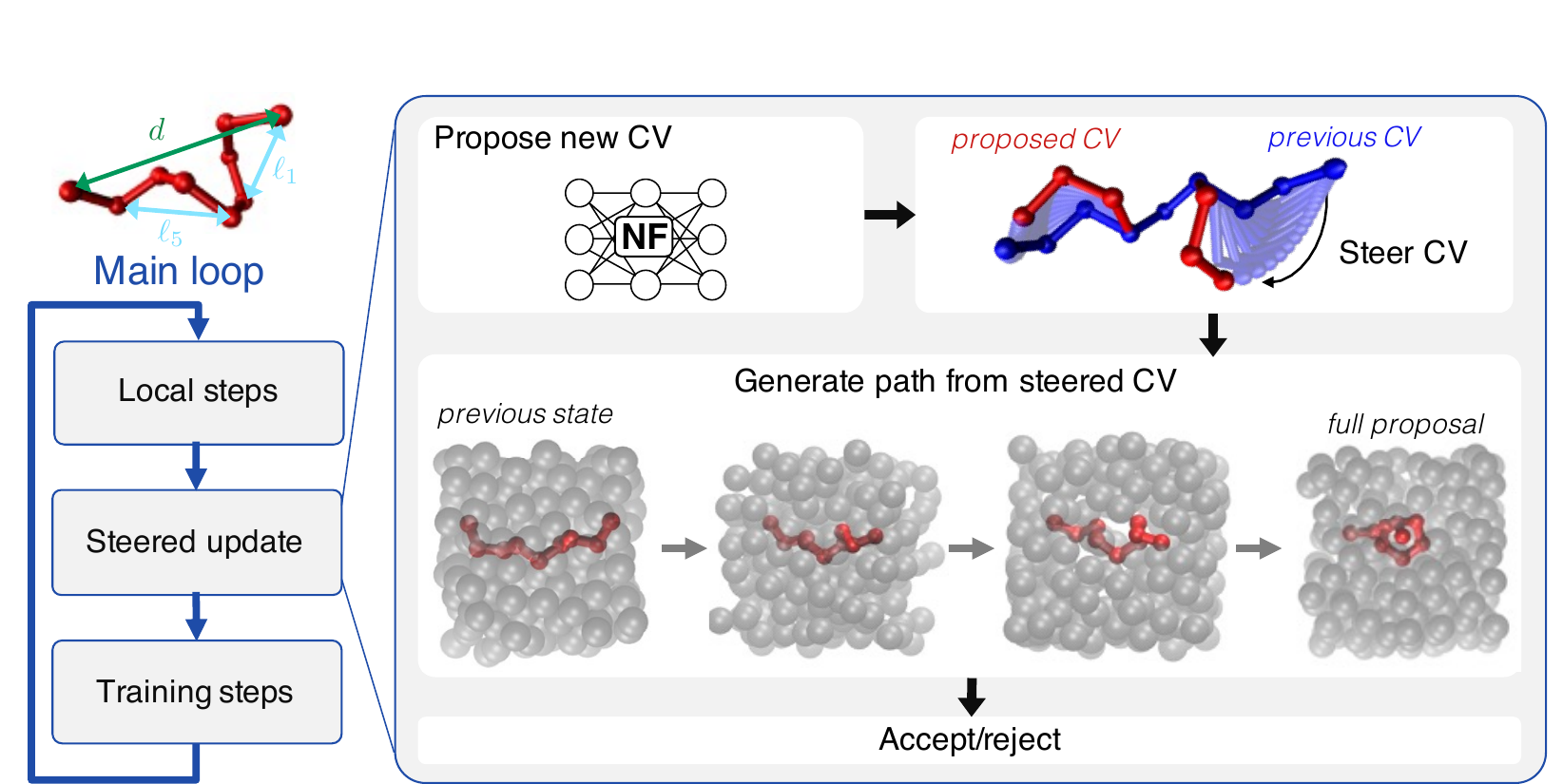}
    \caption{Cartoon representing the functioning of the algorithm. At each iteration, the state of the system is updated through some local steps with a sampler, MALA in our case.
    Then the next step consists in a steered update: the NF proposes a new value of the internal distances, the polymer is then stretched or compacted through steered MD and the solvent is relaxed accordingly. The newly obtained configuration is rejected or accepted according to the work performed and the ratio of NF-proposal probabilities. After the local and non local update of the system, the NF undergoes some training steps.}
    \label{fig:explicative_image_polymer}
\end{figure}

\paragraph{A model polymer in a  Lennard-Jones solvent}
We then benchmarked the algorithm on a more realistic and challenging test system, a polymer of 9 beads immersed in a bath of particles. The interactions considered are: a) A Lennard-Jones (LJ) potential acting among the particles, b) a LJ interaction between polymer and particles, c) a LJ interaction between polymer beads. The chosen LJ radius is always $2^\frac{1}{6}\sigma$ with $\sigma=1$ \revised{and the coupling is chosen to be the same in all three cases with $\epsilon = 1$}. To control the level of metastability we also include in the energy function a double well potential $V(d)$, where $d$ is the end-to-end distance of the polymer (see sketch in \cref{fig:explicative_image_polymer}). The barrier height can be tuned by changing $V(d)$. 
With this setup the polymer has two metastable states: 
a compact one and an extended one, with different weights due to entropic contributions and \revised{significant} solvation effects. \revised{Namely, the solvent bath shifts the free energy of the compact state with respect to the extended state of more than $1K_BT$ (see Supplementary \cref{fig:free_ene_shift_particles}), mainly because the number of solvent particles in the solvation shell differs for the two metastable states.}
We test two settings for the barrier between the two states by changing the barrier in $ V(d)$ from $7\, K_{\rm B}T$ to $28\, K_{\rm B}T$. For both cases,
we estimated the unbiased probability distribution by umbrella sampling using a biased MD trajectory of $10^9$ steps (see Methods). 

We choose as collective variables the set of internal distances of the polymer, namely the distance $\ell_1$ between monomer 1 and monomer 4, the distance $\ell_2$ between monomer 2 and monomer 5 and so on up to $\ell_6$. Therefore we have a 6-dimensional CV space. These distances fully describe the conformation of the polymer, and have the same information content as the dihedral angles, given the bond-lengths and the angles between adjacent monomers are fixed. In \cref{fig:explicative_image_polymer} we present a schematic representation of \cref{alg:CVampling_algo} applied to this test system.

\begin{figure}
    \centering
    \includegraphics[trim={0cm 0cm 0cm 0cm},clip, width=\textwidth]{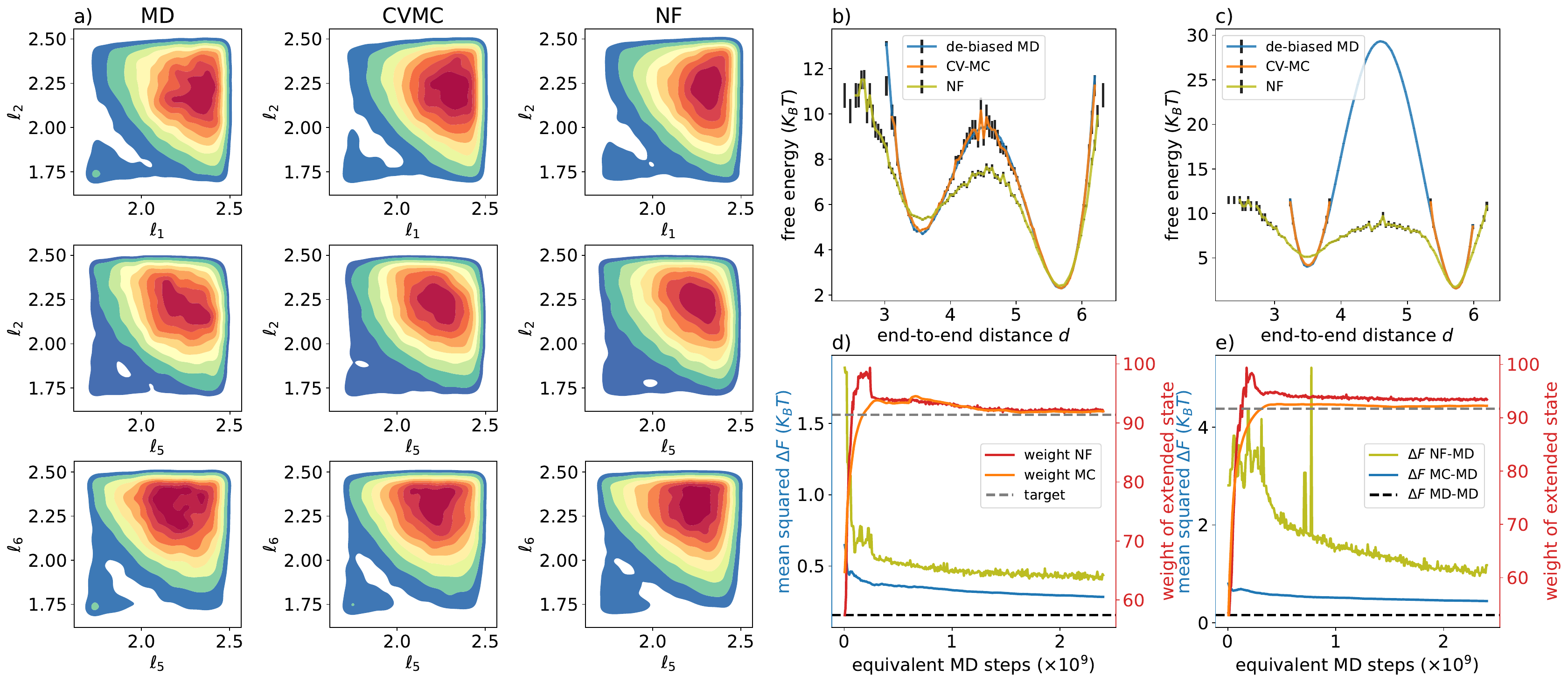}
    \caption{a) 2D slices of the 6D probability density in CV space, with respect to some couples of collective variables. The first column is realized with MD data, the second with the output of \cref{alg:CVampling_algo}, the third with a sample from the trained normalizing flow. b) 1D projection of the free energy on the end-to-end distance of the polymer, in the case where we set a $7\,K_{\rm B}T$ barrier on $V(d)$. The three lines are realized with a biased MD simulation (blue), the output of \cref{alg:CVampling_algo} (orange) and a sample from the trained NF (green). c) 1D projection of the free energy on the end-to-end distance of the polymer, with a $28\,K_{\rm B}T$ barrier on $V(d)$. The three lines are realized with a biased MD trajectory (blue), with the output of \cref{alg:CVampling_algo} (orange) and a sample from the trained NF (green). d) Mean squared error between the 6D free energy obtained from the biased MD simulation and the sample from the trained NF (green line) and the sample from \cref{alg:CVampling_algo} (blue line). The dashed black horizontal line represents the error obtained comparing two distinct MD trajectories. The orange and red line represent the weight of the extended configuration of the polymer with the sample obtained both with the CV-MC (orange) and with NF (red), this is all compared with the target weight (grey dashed line) e) 
    The results are the same as in d), but for the high barrier case.}
    \label{fig:polymer_simulation}
\end{figure}

In \cref{fig:polymer_simulation} we show the results.
As mentioned above, the free energy landscape of this test case is in 6D, with two minima, one corresponding to the compact state, the other to the extended state. These two minima cannot be spotted in any marginal of the probability density as a function of a single CV, but appear on 2D marginals. In panel a) we show some of these 2D marginals for the low barrier case, estimated from an ensemble of 100,000 independent configurations. On each column, the samples were produced in a different way. The first corresponds to the reference MD trajectory, which is obtained by umbrella sampling and subsequent reweighting of the obtained samples. The second one is the visited configurations of \cref{alg:CVampling_algo}, while the third is obtained by sampling in one shot from the NF at the end of the simulation, therefore without the Monte Carlo correction of the visited configurations.
One can notice that MD and CV-MC marginals are more alike than MD and pure NF marginals, but in general there is a good agreement between the three marginals, showing that the algorithm works well.
 
In panel d) and e) we provide a quantitative measure of the difference in the free energies in 6D estimated by the different approach, for the low and high barrier case. 
For this, we used the PAk density estimator from \cite{dadapy} on the CV-MC and MD datasets using the method developed in \cite{mcarli_free_ene_estimate}. For the NF, the normalized density is available analytically.
The figures show the mean squared error (mse) between the free energy estimated by the brute-force MD simulation (considered as ground truth) and the one estimated by NF (green line) and by the CV-MC datasets (blue line), along the iterations of the algorithm.
Since \cref{alg:CVampling_algo} alternates local MALA steps and non local moves with steered MD, each iteration has a cost in terms of what we have called \say{equivalent MD steps}, which is calculated multiplying the number of walkers running in parallel and the number of local steps at each iteration and summing that to the number of walkers times the average number of steps taken by the steered MD step. 
In this way, we can monitor the convergence of the free energy of the NF and CV-MC samples to the ground truth as a function the computational cost.
As a baseline for the expected error on the free energy, we computed the mean squared difference between estimators obtained by two independent biased and reweighted MD trajectories (which could in principle both be considered as ground truth), and reported it as a black dashed line in \cref{fig:polymer_simulation}d). This line can be considered as a lower bound for the best possible result for the NF and CV-MC mse. As shown in the plots of \cref{fig:polymer_simulation}d), the free energy estimated by CV-MC (blue line) converges quickly to the estimate of the MD simulation. This is not the case for the free energy estimated from NF samples (green line). 
In fact after an initial phase of learning, the NF free energy does not improve anymore. This is mainly due to the poor amount of points on the energy barrier, which result in a impossibility for the NF to learn its proper shape, and, possibly, to the insufficient expressiveness of the NF. The red and orange lines represent the  probability mass in the free energy minimum corresponding to the extended state of the polymer obtained with the NF and with the CV-MC samples as a function of the number of the number of equivalent MD steps. 
The reference value obtained with MD is represented as a grey dashed horizontal line. In contrast to what happens with the mse of the 6D free energy, here both the NF and the CV-MC converge to the expected result. This means that the failure of the NF to represent the barrier, and the unnatural connection that it draws between the two free energy minima, causes errors in the free energy landscape, but not in the aggregated relative probability of the two minima. 

In \cref{fig:polymer_simulation}e) we show the results of the same analysis, but relative to the high barrier case. The sample from CV-MC has a free energy that converges to the ground truth with an accuracy compatible with the low barrier case. The NF sample, instead, has a free energy which is much more distant from the true one. Here, there are no samples in the very unlikely region around the barrier, making it impossible to learn for the NF. Yet again, the weight of the extended state, instead, is well represented even in this case (red and orange lines). This shows the power of the procedure, as we can efficiently estimate free energy differences between states separated by a $28\,K_{\rm B}T$ barrier at the same cost as with a $7\,K_{\rm B}T$ barrier.

In \cref{fig:polymer_simulation}b) and c) we show the marginal free energy of the system calculated with respect to the end-to-end distance $d$ of the polymer. Notice that this variable is not part of the CV that we have chosen for the coarse graining, but it is useful to visualize the two distinct minima.
In \cref{fig:polymer_simulation}b) we show the result of the application of \cref{alg:CVampling_algo} to the low barrier case, we compare the result of the output of the CV-MC sampling (orange line) and a sample from the trained NF without MC correction (green line) with the biased MD trajectory (blue line). 
The CV-MC and the MD curves are very well overlapped, showing that the new algorithm is working as expected and can be used to have a good estimate of free energy differences. The NF curve, instead, shows that the barrier is poorly represented and, as a consequence of overall normalization, also the shape of the minima does not perfectly match with the ground truth especially for the higher minimum.
In figure \cref{fig:polymer_simulation}c) we present the same result, but in a case in which the barrier was set to $28 K_{\rm B} T$. Here, we exploit the real power of \cref{alg:CVampling_algo}, as we are able to sample efficiently enough jumps between one minimum and the other even with a very high barrier between them, at the same computational cost as the low barrier case.
The overlapping between the free energy obtained by umbrella sampling and the CV-MC free energy is excellent. For what concerns the NF sample, instead, we face again the same problems as in the low barrier case.

As a final remark we mention that the average acceptance of the non-local NF proposal in this augmented Monte Carlo is greater than 20\% both in the low and in the high barrier case. In the Supplementary Information we show a plot of the average acceptance computed after each resampling step.

\revised{
\paragraph{Reducing the set of CV for the polymer}
The six internal distances of the polymer considered above are enough to specify fully the polymer configuration, because the bond length and the bond angles are fixed.
To show that \cref{alg:CVampling_algo} performs well even with a less informative CV, we show in Supplementary \cref{fig:low_dim_cv} the outcome of a simulation obtained using a 3D collective variable built by choosing only three out of the six internal distances of the polymer. 
In this case the NF 
proposes 3 internal distances such that we cannot reconstruct a complete polymer configuration and end-to-end distance to confront its statistics against MD samples.
However, comparison between the output of \cref{alg:CVampling_algo} and the biased MD trajectory shows as good results as the ones obtained with a 6D CV.
The only difference that we obtain is that the average acceptance of the steered moves is slightly reduced from 20\% to roughly 15\% \cref{fig:low_dim_cv} c). Remarkably this drop does not yield a slow down in convergence time as we found that the non local moves proposed in the 3D CV space 
can be steered with a smaller number of steps with respect to the ones in 6D.}

\section{Discussion}
In this work we introduced an algorithm which exploits a NF to perform enhanced sampling in a coarse-grained space defined by a mid-size CV vector, while generating Boltzmann-distributed configurations of the whole fine-grained system.
We first tested this approach on a mixture of Gaussians, considering only some of the coordinates as CVs. The algorithm has shown a good performance in sampling from this multimodal distribution, even with highly separated modes. The relative weights of the different modes were perfectly reproduced and the marginal distribution on the CV space was correctly learnt by the NF and well represented by the samples generated by the algorithm. This simple test case allowed us to verify that the efficiency is only weakly affected by the energy barrier separating the different modes and by the number of transversal degrees of freedom, if the CV set is properly selected.

We then used the same approach to sample the configurations of a much more complex toy model, a small polymer in a Lennard-Jones bath. This test case added two main complexities to the sampling procedure. First of all, the total number of degrees of freedom passed from being on the order of tens to several hundreds. Secondly, the interaction between the CV and the transversal degrees of freedom became non trivial, since moving the internal distances (or equivalently the dihedrals) of a polymer in a bath causes LJ repulsion between the bath particles and the beads of the polymer that is being moved. These complexities forced us to develop a non trivial scheme to produce the steered moves as in \cref{alg:steered_algo}, by finding an approximate law for the optimal $N(\Vert \Delta\cv\Vert_2)$. 
We have shown that even in this complex case the algorithm works remarkably well both with a low and with a high energy barrier between the two modes of the system, and produces results that are compatible with those of standard methods of simulation.

Our approach offers some important advantages for sampling multimodal distributions in molecular systems. First of all, the use of a NF for making proposals in the CV space allows us to use a large number of collective variables. This is very useful because in most of the enhanced sampling problems choosing the proper minimal set of CV is typically one of the hardest tasks. Moreover the efficiency of the present enhanced sampling techniques is high only when the CV space is low dimensional, while with the NF we can afford to include tens of collective variables. Another very important advantage of our approach comes from the fact that, if the CV set is properly chosen, the acceptance of a move between two points does not depend on the height of the free energy barrier between them. 
In the case of the polymer, the dependence of the average acceptance on the barrier is very weak even though there is a non trivial coupling between transversal degrees of freedom and CV, confirming again the validity of this result. One more important advantage of the algorithm is the good scaling that it shows with the number of transversal degrees of freedom, this is proven by the Gaussian mixture example and by the large number of solvent particles used in the polymer toy model. Finally, one great advantage is that, once trained, the NF can be used at will to have a very efficient Monte Carlo to produce rapidly uncorrelated configurations for the system of interest. 

\revised{We also note that the proposed algorithm shows promises for systems in which adequate CVs are not known a priori. Firstly, the ability of NFs to model hundreds of dimensions allows to envision greedy approaches where a large fraction of easily identifiable relevant degrees of freedom, such as atoms of the main chain of a biomolecule, are kept as CVs. Secondly, the iterative nature of the procedure refining sequentially the sample of configurations and the NF model learned from this sample, could further include the parallel learning of a CV using recently proposed data-based inference methods, often relying also on deep learning \cite{bonatiDeepLearningSlow2021,belkacemiChasingCollectiveVariables2022}.}

The procedure still shows weak points that will have to be improved in a future work. One downside is given by the fact that the NF is not able to perfectly learn the underlying marginal probability density function, as one can see for instance from \cref{fig:polymer_simulation}b) to e). This causes some small biases in the CV proposal distribution that on one side affect the overall acceptance, on the other side do not allow us to use directly the trained NF to produce samples without the need of a MC step to re-weight them. A more important drawback is that moves that require travelling a large distance in CV space have a good acceptance only if performed using a large number of MD steps during the steering (see \cref{fig:polymer_nsteps_acc} a)). This affects the overall efficiency of the algorithm, making it competitive with other enhanced sampling methods only when we have to overcome large free energy barriers between different minima, or when we cannot choose a small set of collective variables. \revised{As such, the steps to be taken to learn the free energy of a protein in water are many, including merging the framework into a MD code.}

For future developments, we will need to improve how the chosen map reproduces the underlying marginal probability density function of the system. This improvement can be addressed in two directions. On one side, we could test the performance of more sophisticated generative models in the CV space, e.g. using diffusion models and working directly in Cartesian coordinates with equivariant models \cite{pmlr-v202-yim23a,midgleySEEquivariantAugmented2023a,klein2023timewarp,akhound-sadeghIteratedDenoisingEnergy2024}. On the other side, we could improve the chosen loss function, and make it more suitable for learning better also the shape of the free energy barrier.

\section{Material \& Methods}

\paragraph{Mixture of Gaussians}
Let $x=(\cv, x_\perp)$ and $g(x) = \mathcal{N}\left(x-\mu, \Sigma\right)$ be a multivariate Gaussian with mean $\mu \in \mathbb{R}^n$ and covariance $\Sigma \in \mathbb{R}^{n\times n}$, the tested distributions were built as follows: 
\begin{align}
p(x) = \left(w_1 \mathcal{N}\left(\cv - \psi_1, \Sigma_1\right) + w_2 \mathcal{N}\left(\cv - \psi_2, \Sigma_2\right) \right) \times \mathcal{N}\left(x_\perp - \vec{0}, \mathbb{I} \right)
\end{align}
In all the tests we considered $\cv \in \mathbb{R}^2$, $w_1=\frac{1}{4}$, $w_2 = \frac{3}{4}$, $\Sigma_1$ has off-diagonal elements $\sigma_{12} = \sigma_{21}= -3.5 \times 10^{-2}$ and diagonal elements $\sigma_{ii}=5\times 10^{-2}$, $\Sigma_2$ is diagonal with $\sigma_{ii}=0.2$. For \cref{fig:gaussian_experiment}a) to f) we have considered only one transversal degree of freedom and $\cv_1 = (-m, m)$, $\cv_2 = (m, m)$ with $m=1.84$. For \cref{fig:gaussian_experiment}g) we have used one transversal degree of freedom but varied $m$ in the set $\{1.0, 1.21, 1.42, 1.63, 1.84, 2.05, 2.26, 2.47, 2.68, 2.89\}$. 
Since there are two modes with different weights, the free energy has two minima, one lower than the other and a single maximum in between. The energy barrier is calculated taking the difference between the maximum and the highest minimum. 
For \cref{fig:gaussian_experiment}h) we kept $m=1.84$ and changed the number of transversal degrees of freedom from $1$ to $28$.

In all simulations, the local steps were performed using MALA as described in \cref{alg:mala_algo} with parameters $\beta = 1$, friction $\gamma = 1$, and time-step $\tau = 0.005$. The steered moves had $N=20
$ steps and 60 independent random walkers were initialized in each minimum. 

The architectures of the tested NFs were as follows: for the RealNVP we have used 
40 affine coupling layers where the scaling and translation networks were multilayer perceptrons with 3 hidden layers, all with width equal to 10.
The base distribution was chosen to be a Gaussian with average and covariance taken from a population of independent walkers evenly distributed among the minima and evolved for some MC steps inside the minima to have them decorrelated. For the RealNVP, it was essential to have a number of training iterations on the order of tens of thousands as the convergence of the loss function was relatively slow. We used  40000 training iterations with learning rate 0.0025. For the RQS, we leveraged the library \texttt{nflows}\cite{nflows} and used splines with 10 bins between $-5$ and 5 and 3 layers, the parameters of the spline were obtained with multilayer perceptrons of depth 6 and 12 hidden features per layer. The base distribution, in this case, was a standard normal. The RQS not only has a much better representational power, but also converges faster, in about 2000 iterations, using the same learning rate as the one for the RealNVP.

\paragraph{The polymer in a Lennard-Jones solvent: MD simulation}
The simulation was performed using MALA as described in SI \cref{alg:mala_algo}. For the interaction between different particles we have used a Lennard Jones potential: $U_{\rm LJ}(r)= 4\epsilon [ \left( \frac{\sigma}{r}\right)^{12} - \left( \frac{\sigma}{r} \right)^6 ]$. We have chosen for the coupling interaction $\epsilon_{ss}=\epsilon_{pp}=\epsilon_{ps}=1$, for solvent-solvent, polymer-polymer and polymer-solvent interaction respectively, and we have set $\sigma=1$ in all the cases. All the energy units were then set relative to $\epsilon_{ss}$. As parameters for the dynamics we have chosen $K_{\rm B}T = 1.3\, \epsilon_{ss}$, $\gamma = 1$, $\tau = 0.0002$. The time step $\tau$ was chosen in such a way to have the fastest possible decorrelation among successive MALA steps, which occurs when the average MALA acceptance is roughly 70\%. The boundary conditions were chosen to be periodic, in a cubic box of size $7.5$ (all the lengths are relative to the polymer bond length, which is chosen to be 1). The potential on the end-to-end distance has the form $V(d)=\frac{v}{4}\left(\frac{d-d_0}{\alpha}\right)^4-v\left(\frac{d-d_0}{\alpha}\right)^2$, with $d_0=4.62$ and $\alpha=0.8$. We set $v=9.1$ or $v=36.4$ yield respectively $\sim 7\,K_{\rm B}T$ and $\sim 28\,K_{\rm B}T$ barriers between the extended and the compact state. 
The particles are evolved with an overdamped Langevin dynamics, while the polymer is evolved through the integration of a second order Newtonian dynamics, with a chosen monomer mass $m=0.1$. There is no thermostat linked to the polymer, as it already thermalizes with the surrounding bath of LJ particles. After every step of the system evolution we set a Metropolis acceptance rule. The dynamics of the polymer is constrained: the bond lengths are all fixed to 1 and all the angles between adjacent monomers are fixed to $70.5^\circ$. Moreover the frame of reference of the simulation is chosen in such a way that the central monomer of the polymer is fixed, therefore the only relevant degrees of freedom of the polymer are the dihedral angles or, equivalently, the internal distances, as there is a one to one mapping between the two. For this reason, all the forces are projected on the internal distances and the dynamics is done directly on them. Even though the dynamics of the polymer is calculated directly on the internal distances, at each iteration, we need to know the 3D coordinates of every monomer to be able to calculate the forces coming from the LJ interactions. To do so, we use the NeRF algorithm \cite{nerfalgorithm} to pass from the internal coordinates of the polymer to the Cartesian ones. To ease the computation of forces we compute derivatives with the autodifferentiation library Zygote \cite{zygotelibrary}. The reference MD trajectory used in the analysis in \cref{fig:polymer_simulation}b) and c) was obtained with $10^9$ steps, saving one configuration each $10^4$ steps. To obtain the free energy both in the high and low barrier case we used umbrella sampling, with a bias potential $V_B(d)= - V(d)$ (which corresponds to switching off the potential on the end-to-end distance).

\paragraph{The polymer in a Lennard-Jones solvent: CV-MC simulation}
The simulation was performed with 200 random walkers running in parallel, half of them initialized in the compact state and half initialized in the extended state.
The local moves corresponded to 30 local MD updates with the schedule explained above at each iteration of \cref{alg:CVampling_algo}.

For setting up properly the non local moves we have done some preliminary study on the system.

We first estimated the optimal number of steered steps $N(\cdot)$ for any proposed steered update, as a function of the size of the CV jump $\Vert\Delta\cv\Vert_2$.
$N(\Vert\Delta\cv\Vert_2)$ is the value that maximises the ratio between the probability of acceptance $p$ \eqref{eq:acc-prob} over the computational cost of the steered update, which is given by the number of intermediate steps $N$ that we take in a steered move. We consider a for a set of values $\Vert\Delta\cv\Vert_2$ spanning well the CV space, and for each estimate $N(\Vert\Delta\cv\Vert_2$) from a set of independent transitions between multiple pairs of points $(\cv,\cv')$ that have the same distance $\Vert \cv - \cv'\Vert_2$ in CV space. 
More precisely, we consider for this study loop transitions: from $\cv$ value to $\cv'$ and back to $\cv$. Thanks to this construction the free energy difference of the loop transition is zero, and the acceptance of the transition is influenced only by the work performed on the system, it is also independent of the direction of the transition $\cv \to \cv'$ versus $\cv' \to \cv$.
The results of the optimal number of steps in the steered update for a given $\Vert \cv - \cv'\Vert_2$ are presented in the SI \cref{fig:polymer_nsteps_acc}a). There are two main regimes with a sharp transition. The first regime, where the optimal number of steps is on the order of a few hundreds, corresponds to small scale transitions where the NF proposes new configurations inside the same mode, namely transitions from compact state to compact state or from extended to extended. The second one, where the optimal number of steps is on the order of tens of thousands corresponds to large scale transitions, from one stable state to the other were the polymer is stretched or compacted and all the particles around it have to adapt to the new conformations.

To accelerate the learning of the NF, we have pretrained the network on configurations obtained by local MD simulations with an equal number of walkers in the two minima of the potential.  
The NF was then learns first the presence of two modes but considering their weights are equal. This explains why the acceptance of the resampling steps is quite high even at the beginning of the algorithm iterations, as one can see from \cref{fig:polymer_nsteps_acc}b).
We use the same RQS architecture as in the Gaussian example. After the pretraining phase, the full algorithm was run using as initialization for the NF the pretrained RQS, the learning rate was set to $0.0045$, the total number of training iterations was $12\,000$, and the total number of resampling steps was 400.

\section*{Acknowledgements}

We thank Tony Lelièvre, Christoph Sch\"onle, Pilar Cossio, Gabriel Stoltz and Eric Vanden-Eijnden for insightful discussions. The work of M.G. is supported by the Hi! Paris center.
Work partially funded by NextGenerationEU through the Italian National Centre for HPC, Big Data, and Quantum Computing (grant number CN00000013).

\bibliographystyle{alpha}
\bibliography{references}

\pagebreak
\appendix
\section*{Supplementary information}

%% Add an S prefix to the numbering in the supplementary
\makeatletter
\renewcommand{\thefigure}{S\@arabic\c@figure}
\makeatletter
\renewcommand\thealgorithm{S\arabic{algorithm}}

\begin{algorithm}[h!]
\caption{Metropolis-Adjusted Langevin Algorithm (MALA)}\label{alg:mala_algo}
\begin{algorithmic}
\Require $U(\cdot)$ system energy function, $\beta$ inverse temperature, $\tau$ integration time step, $\gamma$ friction coefficient, $x(0)$ starting configuration, $n_{\rm steps}$ number of total steps.
\\
\For{$n = 1 \dots n_{\rm steps}$}
\State $\widetilde x = x\left(n\right) - \frac{\tau}{\gamma}\nabla U\left(x\left(n\right)\right) + \sqrt{\frac{2\tau}{\beta\gamma}}\eta, \quad \eta \sim N\left(0,1\right)$ \Comment{propose new configuration with overdamped Langevin step}
\State $\Tilde{\eta} = \sqrt{\frac{\beta\tau}{2\gamma}}\left(\nabla U\left(x\left(n\right)\right) + \nabla U\left(\widetilde x\right) \right) - \eta$
\State $\log R = -\frac{1}{2}\left(\Tilde{\eta}^2-\eta^2\right) - \beta\left(U\left(\widetilde x\right)-U\left(x\left(n\right)\right)\right)$ \Comment{calculate acceptance probability}
\State $u \sim \rm \mathcal{U}([0,1])$\Comment{where $\mathcal{U}$ is the uniform law.}
\If{$u<\min\{1,R\}$} \Comment{accept or reject the proposal}
\State $x\left(n+1\right) = \widetilde x$ 
\Else
\State $x\left(n+1\right) = x\left(n\right)$
\EndIf
\EndFor
\\
\Return $\{x\left(n\right)\}_{n=1, ..., n_{\rm steps}}$
\end{algorithmic}
\end{algorithm}

\begin{algorithm}[h!]
\caption{Steered Update}\label{alg:steered_algo}
\begin{algorithmic}
\Require $U(\cdot)$ system energy function, $\beta$ inverse temperature, $\tau$ integration time step, $\gamma$ friction coefficient, $x=(\cv, x_\perp)$ starting configuration, $N(\cdot)$ steering schedule discretization rule, $\rho_\theta^\Psi(\cv)$ proposal probability parametrized by the flow. 
\\
\State $(\cv^0, x_{\perp}^0) = (\cv, x_\perp)$ \Comment{store starting point}
\State $\cv' \sim \rho_\theta^\Psi(\cv)$ \Comment{draw new set of collective variables}
% \State $n_{\rm steps} = N(\Vert \cv'-\cv\Vert_2)$ \Comment{define schedule}
\State $\Delta\cv = \frac{\cv'-\cv}{2N(\Vert \cv'-\cv\Vert_2)}$
\State $\Delta S = 0$ \Comment{initialize action produced}
\While{$n < N(\Vert \cv'-\cv\Vert_2)$}
\State $\cv \gets \cv + \Delta\cv$ \Comment{update CV}
\State $x_\perp \gets {\rm MALA}(U(\cv, \cdot), \beta, \gamma, \tau, x_\perp, n_{\rm steps}=1)$ \Comment{update transversal d.o.f. with 1 MALA step at fixed $\cv$}
\State $\Delta S \gets \Delta S + U(\cv, x_\perp)-U(\cv, x_\perp')$ \Comment{save the change in energy}
\State $\cv \gets \cv + \Delta\cv$ \Comment{update CV}
\State $n \gets n+1$
\EndWhile
\State $R = \frac{\rho_\theta^\Psi(\cv^0)}{\rho_\theta^\Psi(\cv)} \exp\left({-\beta(\Delta S + U(x_\perp, \cv)-U(x_{\perp}^0, \cv^0))}\right)$ \Comment{calculate acceptance probability}
\State $u \sim \rm \mathcal{U}([0,1])$\Comment{where $\mathcal{U}$ is the uniform law}
\If{$u<\min\{1,R\}$} \Comment{accept or reject the proposal}
\\
$\quad$\Return $(\cv, x_\perp)$
\Else
\\
$\quad$\Return $(\cv^0, x_{\perp}^0)$
\EndIf
\end{algorithmic}
\end{algorithm}

\begin{figure}[h!]
    \centering
    \includegraphics[trim={0cm 0cm 0cm 0cm},clip, width=\textwidth]{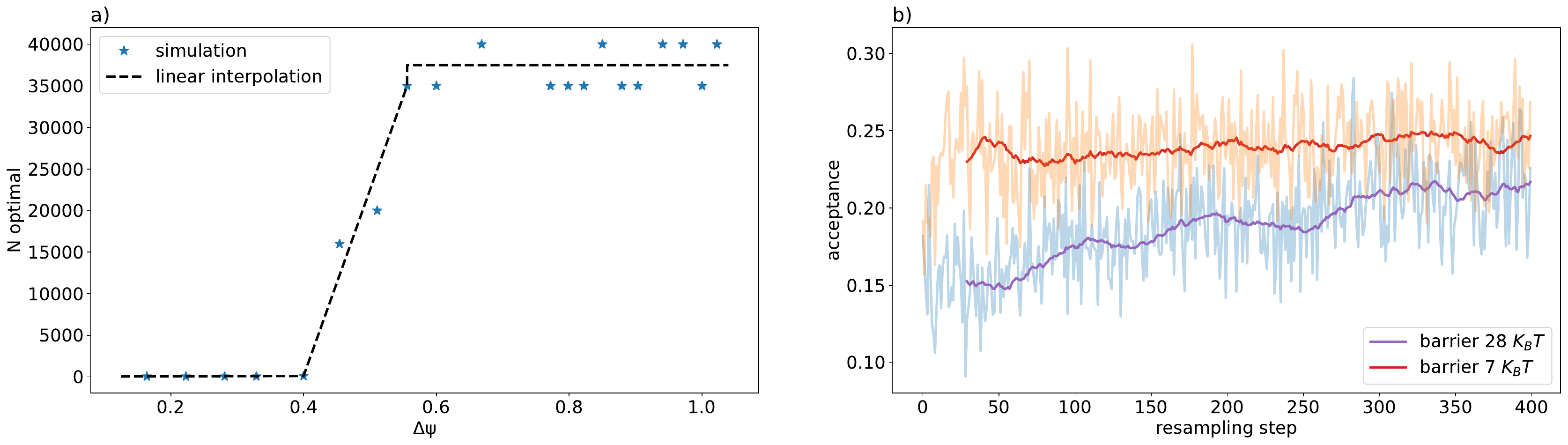}
    \caption{a) Optimal number of steps for a steered transition as a function of the distance $\Vert \Delta\cv\Vert_2$ travelled in CV space. b) Acceptance of the non-local steered moves (resampling steps) averaged among the 200 parallel random walkers for the two tried barriers.}
    \label{fig:polymer_nsteps_acc}
\end{figure}

\begin{figure}[h!]
    \centering
    \includegraphics[trim={0cm 0cm 0cm 0cm},clip, width=0.5\textwidth]{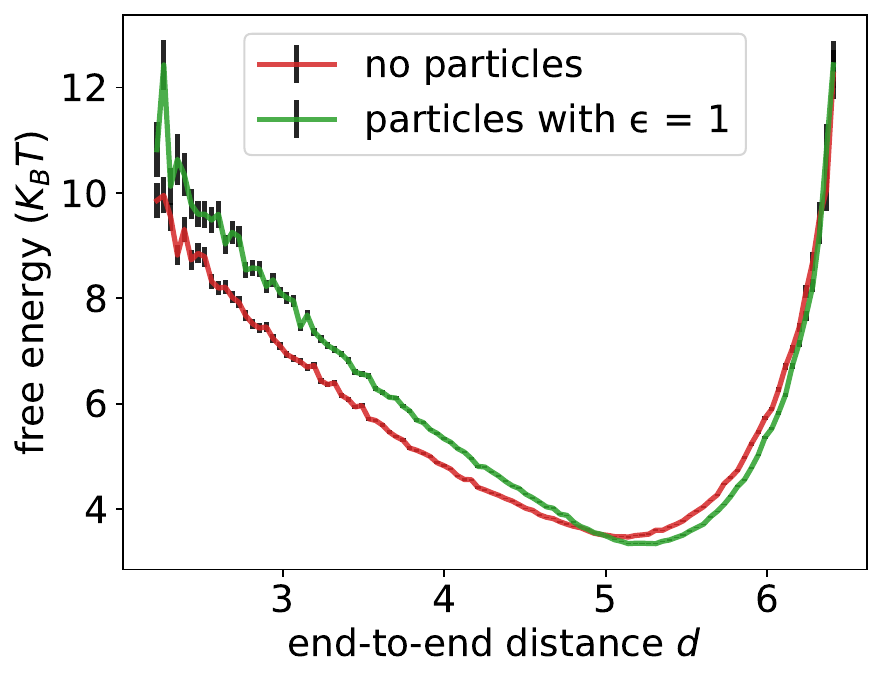}
    \caption{Free energy profile with respect to the end to end distance of a biased trajectory for a polymer in a Lennard-Jones bath with coupling $\epsilon = 1$ (green line) and for a polymer coupled with a thermostat, but without a bath of particles (red line), the total free energy shift is more than $1 K_BT$. }
    \label{fig:free_ene_shift_particles}
\end{figure}

\begin{figure}[h!]
    \centering
    \includegraphics[trim={0cm 0cm 0cm 0cm},clip, width=\textwidth]{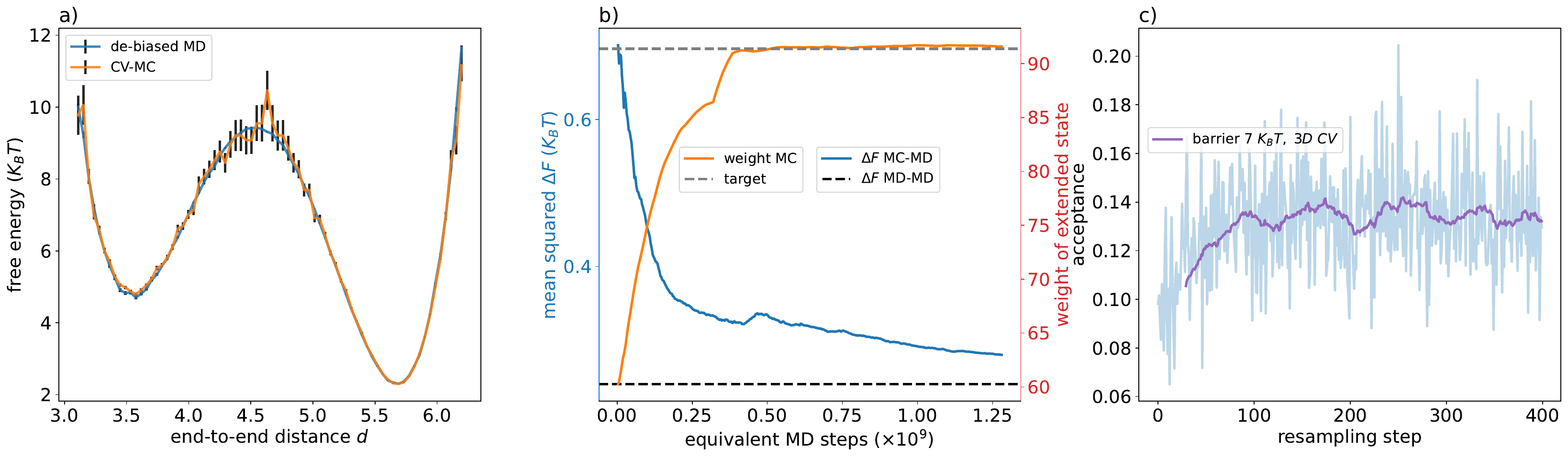}
    \caption{a) Free energy profile with respect to the end to end distance for a polymer simulation performed using only 3 internal distances as collective variable (orange), confronted with the ground truth obtained with biased Molecular Dynamics (blue). b) Convergence of the relative weight (orange) of the modes of the probability distribution obtained with the simulation with a 3D collective variable to the ground truth weight, and convergence of the mean squared error between the free energy obtained with the collective variable Monte Carlo and the ground truth one (blue). c) Acceptance of the non-local steered moves (resampling steps) averaged among the 200 parallel random walkers.}
    \label{fig:low_dim_cv}
\end{figure}

\end{document}